\documentclass[11pt,a4paper]{article}
\usepackage{jcappub}

\usepackage{array,calc,epsfig}
\usepackage{bbm}
\usepackage{xcolor}

\newcommand{\md}{\mathrm{d}}
\def\be{\begin{equation}}
\def\ee{\end{equation}}
\def\bseq{\begin{subequations}}
\def\eseq{\end{subequations}}

\def\bea{\begin{eqnarray}}
\def\eea{\end{eqnarray}}

\def\bseq{\begin{subequations}}
\def\eseq{\end{subequations}}

\title{Cuscuton Inflation}
\author[a,b,c]{Nicola Bartolo,}
\author[d]{Alexander Ganz,}
\author[a,b,c,e]{Sabino Matarrese}

\affiliation[a]{ Dipartimento di Fisica e Astronomia ``G. Galilei", \\
	Universit\`a degli Studi di Padova, via Marzolo 8, I-35131 Padova, Italy}
\affiliation[b]{INFN, Sezione di Padova, \\ via Marzolo 8, I-35131 Padova, Italy}
\affiliation[c]{INAF - Osservatorio Astronomico di Padova, \\ Vicolo dell’Osservatorio 5, I-35122 Padova, Italy}
\affiliation[d]{Faculty of Physics, Astronomy and Applied Computer Science, Jagiellonian University, 30-348 Krakow, Poland}
\affiliation[e]{Gran Sasso Science Institute, \\ Viale F. Crispi 7, I-67100 L'Aquila, Italy}

\emailAdd{nicola.bartolo@pd.infn.it}
\emailAdd{alexander.ganz@uj.edu.pl}
\emailAdd{sabino.matarrese@pd.infn.it} 

\abstract{We study the impact of (generalized) cuscuton models on standard single scalar field inflation. Generalized cuscuton models are characterized by spatial covariant gravity where a scalar degree of freedom is made non dynamical, and there are just two tensor degrees of freedom. The presence of the non-dynamical scalar field does not spoil inflation but instead the modifications are, in general, slow-roll suppressed leading to almost scale-invariant power spectra.  However, the extra free parameters, which can be tuned relatively independently, lead to a larger parameter range for observable quantities, such as the tensor-to-scalar ratio.    
For the (generalized) cuscuton model the non-Gaussianties of the curvature bispectrum are suppressed by the slow-roll parameters, and, therefore, outside the reach of current experiments. However, generalized cuscuton models can lead to a different shape for the bispectrum which might be constrained by future  experiments.
}

\arxivnumber{}

\notoc 

\begin{document}

\maketitle
\flushbottom

\pagebreak
\section{Introduction}
General relativity (GR) is an extremely successful theory of gravity, however, since the detection of an accelerated expansion of the Universe modified theories of gravity has obtained an increased attention. The Lovelock theorem \cite{doi:10.1063/1.1665613,doi:10.1063/1.1666069} offers a guideline how to modify gravity stating that GR with a cosmological constant is the most general theory in a four dimensional spacetime with second order equations of motion respecting general covariance and locality and written only in terms of the metric. 
Therefore, a conventional way to extend GR is by adding additional degrees of freedom. However, in recent years there has been a growing attention to modification of GR without extra degrees of freedom by breaking the spacetime diffeomorphism invariance \cite{Iyonaga:2018vnu,Gao:2019twq,Lin:2017oow,Mukohyama:2019unx,Yao:2020tur,Lin:2020nro,Ganz:2021hmp}. 

The first approach has been developed in the cuscuton model \cite{Afshordi:2006ad}, a specific kind of scalar-tensor theories. In the unitary gauge the extra degree of freedom becomes non-dynamical leading to a theory of gravity with just two tensor degrees of freedom \cite{Gomes:2017tzd}. This feature is similar to the existence of the preferred foliation in the context of the Horava-Lifshitz theories \cite{Blas:2009yd} or higher order scalar-tensor theories \cite{Crisostomi:2017ugk,DeFelice:2018ewo,Ganz:2019vre}, where the unitary gauge prevents the existence of additional Ostrogradsky ghosts. While working in the unitary gauge may be risky since the scalar degree of freedom can reappear around non-homogeneous backgrounds in \cite{DeFelice:2018ewo,DeFelice:2021hps} it has been argued that these modes are non-physical and can be removed by imposing proper boundary conditions.

Although the cuscuton model does not contain extra degrees of freedom in the unitary gauge it provides interesting new features as a distinct Cosmic Microwave Background \cite{Afshordi:2007yx}, stable bouncing solutions \cite{Boruah:2018pvq,Quintin:2019orx,Kim:2020iwq}, viable power-law solutions of inflation \cite{Ito:2019fie}, an accelerated universe with an extra dimension \cite{Ito:2019ztb} and the absence of caustic instabilities which are a common problem of scalar tensor theories \cite{deRham:2016ged}. Further, it has been discussed that the cuscuton model with a quadratic potential can be regarded as a low-energy limit of the non-projectable Horava-Lifshitz gravity model \cite{Afshordi:2009tt,Bhattacharyya:2016mah}. 

In this work we want to discuss the impact of the (generalized) cuscuton model on canonical single scalar field inflation. We will show that the presence of the non-dynamical scalar field does not spoil inflation consistent with previous results \cite{Ito:2019fie}. Indeed at the linear level we recover an almost scale invariant power spectrum of curvature and tensor perturbations and the corrections coming from the (generalized) cuscuton are proportional to the slow-roll parameters. Further, we will discuss higher order perturbation effects by analyzing the bispectrum of scalar perturbations. While in the cuscuton models the non-Gaussianties similar to the canonical single scalar field are outside the reach of current experimental constraints \cite{Planck:2019kim}, generalized cuscuton models can lead to a different shape of the bispectrum  which might be constrained in future experiments.  The structure of our work is split into two parts. In section \ref{sec:Cuscuton_Inflation} we will discuss the impact of the cuscuton field on single scalar field inflation. First, we will recap the linear power spectra generalizing the work of \cite{Ito:2019fie} and then discuss the implications for the bispectra.
In the second part, section \ref{sec:Generalized_Cuscuton}, we will extend our analysis to generalized cuscuton-like models introduced in \cite{Gao:2019twq}. We will focus on the main generic features for models inside or outside the  Gleyzes-Langlois-Piazza-Vernizzi (GLPV) class also known as beyond Horndeski \cite{Gleyzes:2014dya}.
In section \ref{sec:Conclusion} we end with a conclusion of our results.

In our paper we are using the mostly plus signature $(-,+, +, +)$ for the metric. Greek indices are running from 0 to 4 and Latin ones from 1 to 3. Further, we are using units in which the speed of light and the reduced Planck mass are set to one.

\section{Cuscuton inflation}
\label{sec:Cuscuton_Inflation}
We analyze the single scalar field inflation model in the presence of the cuscuton field. The action is given by
\begin{align}
   S = \int \md^4x\,\sqrt{-g} \left( \frac{1}{2} R + \mu^2 \sqrt{- \partial_\mu \varphi \partial^\mu \varphi} - V(\varphi) - \frac{1}{2} \partial_\mu \chi \partial^\mu \chi - U(\chi) \right) ,
\end{align}
where $\varphi$ is the cuscuton and $\chi$ the standard inflaton field. 
If the cuscuton field is homogeneous $\partial_i \varphi=0$ the scalar field becomes non-dynamical at the full non-linear level leading to a theory of gravity with just two degrees of freedom \cite{Gomes:2017tzd}. 
However, for generic scalar field profiles there is, in general, an additional instantaneous mode (see the discussion in the appendix \ref{app:Instantaneous_modes}).
To avoid issues with potential instantaneous modes coming from a non-homogeneous cuscuton field, we will fix from now on $\varphi=\varphi(t)$ already strongly at the action level. 

Using the standard ADM decomposition of the metric
\begin{align}
    \md s^2 = - N^2 \md t^2 + h_{ij} \left( \md x^i + N^i \md t \right) \left( \md x^j + N^j \md t \right),
\end{align}
where $N$ is the lapse, $N^j$ the shift vector and $h_{ij}$ the 3-dim metric on the hypersurface, the action can be rewritten as
\begin{align}
 S = \int \md^4x\,N \sqrt{h} \big[ & \frac{1}{2} \left( K_{ij} K^{ij} - K^2 + \bar R \right)  - V(\varphi) + \frac{1}{2} \nabla_n \chi \nabla_n \chi \nonumber \\
 & - \frac{1}{2} h^{ij} \partial_i \chi \partial_j \chi - U(\chi) \big] + \mu^2 \int \md^4x\,\sqrt{h} \vert \dot \varphi \vert ,
\end{align}
where we have defined $ \nabla_n \chi \equiv (\dot \chi - N^i \partial_i \chi)/N$, $\bar R$ is the intrinsic curvature and $K_{ij}$  the extrinsic curvature
\begin{align}
    K_{ij} = \frac{1}{2 N} \left( \dot h_{ij} - D_i N_j - D_j N_i \right).
\end{align}

\subsection{Background equation of motion}
Using the Friedmann-Lemaitre-Robertson-Walker (FLRW) metric
\begin{align}
    \md s^2 = - \md t^2 + a^2(t) \delta_{ij} \md x^i \md x^j
\end{align}
the background equations of motion (EOMs) are given by
\begin{align}
    3 H^2 = V + U + \frac{1}{2} \dot\chi^2, \\
    2 \dot H = - \mu^2 \vert \dot \varphi \vert - \dot \chi^2, \\
    \mathrm{sign}(\dot \varphi)\, 3 \mu^2 H + V_\varphi = 0, \\
    \ddot \chi + 3 H \dot \chi + U_\chi =0,
\end{align}
where $H=\dot a/a$ is the Hubble parameter.
For later purposes it is useful to define the slow-roll parameters
\begin{align}
    \epsilon= - \frac{\dot H}{H^2}, \quad \alpha = \frac{\dot \chi^2}{2 H^2}, \quad  \eta = \frac{\dot \epsilon}{H \epsilon}, \quad \beta= \frac{\dot \alpha}{\alpha H}.
\end{align}
Further, let us define 
\begin{align}
\label{eq:sigma}
    \sigma = \epsilon-\alpha = \frac{\mu^2}{2 H^2 } \vert \dot \varphi \vert \ge 0.
\end{align}
Note, that in standard single scalar field inflation $\sigma=0$. Therefore, the parameter indicates the impact of the cuscuton scalar field \cite{Boruah:2017tvg,Ito:2019fie}. 

A specific class of solutions of the background are power-law solutions $a \sim t^p$ \cite{PhysRevD.32.1316} which provide the appealing feature that the EOMs are exactly solvable. In that case we have $\eta = \beta =0$. 
In \cite{Ito:2019fie} it has been shown that it is possible to construct new  power-law solutions for inflation with $\sigma \neq 0$. 

\subsection{Linear perturbations}
\label{sec:Linear_perturbations_Cuscuton}
The linear perturbations of the cuscuton model in the presence of a minimally coupled scalar field has been extensively discussed in \cite{Boruah:2017tvg}. We will just mention the most important steps and refer otherwise to the aforementioned paper for a detailed discussion. 

In the unitary gauge of the cuscuton field
\begin{align}
    h_{ij} = a^2 e^{2 \xi } \left( \delta_{ij} + \gamma_{ij} + \frac{1}{2} \gamma_{ik} \gamma^{k}_j + ... \right), \\
    N = 1 + \delta N, \quad N_i = \partial_i \psi, \quad \chi = \bar \chi + \delta \chi
\end{align}
the second order action for scalar perturbations can be expressed as
\begin{align}
    \delta_2 S = \int \md^4x\,a^3 \Big[ & - 3 \dot \xi^2 + \frac{(\partial_k \xi)^2}{a^2} - 3 H^2 \delta N^2 - 2 H \delta N \frac{\partial^2 \psi}{a^2} + 2 \dot \xi \frac{\partial^2 \psi}{a^2} \nonumber \\ &  + 6 H \delta N \dot \xi - 2 \delta N \frac{\partial^2 \xi}{a^2} - \frac{1}{2} \frac{(\partial_k \delta \chi)^2}{a^2} + \dot \chi \frac{\partial^2 \psi}{a^2} \delta \chi + \frac{1}{2} \dot \chi^2 \delta N^2 \nonumber \\
    & - \dot \chi \delta N \delta \dot \chi +3 \dot \chi \xi \delta \dot \chi + \frac{1}{2} \delta \dot \chi^2 - U_\chi \delta \chi ( \delta N + 3 \xi) - \frac{1}{2} U_{\chi\chi} \delta \chi^2 \Big].
\end{align}
For the ease of notation we have dropped the bar for the background value of the inflaton field $\chi$.
Solving the momentum and hamiltonian constraint yields
\begin{align}
    \delta N = \frac{\dot \xi}{H} + \frac{\dot \chi}{2 H } \delta \chi, \qquad \psi= - \frac{\xi}{H} + \frac{a^2}{\partial^2} \left( \alpha \dot \xi - \frac{\dot\chi}{2H} \delta \dot \chi + \frac{1}{2} \dot\chi  \left( \frac{1}{2} \beta - \sigma \right)  \delta \chi \right).
\end{align}
Redefining $\xi$ as
\begin{align}
\label{eq:Redefining_curvature}
    \tilde \xi = \xi - \frac{H}{\dot \chi} \delta \chi
\end{align}
and substituting the expression for the lapse and shift parameter we obtain that the perturbations of the inflaton are non-dynamical
\begin{align}
\label{eq:Second_order_action_Inflation_Integrating_out}
    \delta_2 S = \int \md^4x\,a^3 \Big[ & \alpha \dot{\tilde\xi}^2- \sigma \dot \chi \dot{\tilde\xi} \delta \chi - \frac{1}{2} \delta\chi^2 H^2 (3\sigma - \sigma^2)  - \sigma \frac{H^2}{\dot\chi^2} \frac{(\partial_i \delta\chi)^2}{a^2} \nonumber \\
    & - \epsilon \frac{(\partial_i \tilde\xi)^2}{a^2} - 2 \sigma \frac{H}{\dot\chi} \frac{\partial_i \tilde\xi \partial_i \delta\chi}{a^2}\Big].
\end{align}
Solving the EOM for the inflaton field $\delta \chi$ we get
\begin{align}
    \delta \chi =& - \frac{2 \alpha \frac{H}{\dot \chi} \frac{\partial^2}{a^2}}{\frac{\partial^2}{a^2} - H^2 \alpha (3-\sigma)} \tilde \xi + \frac{\alpha \dot \chi}{ \frac{\partial^2}{a^2} -   H^2 \alpha (3-\sigma)} \dot{\tilde \xi}.
    \label{eq:solution_chi_linear}
\end{align}
Switching to conformal time we can express the second order action finally as
\begin{align}
    \delta_2 S = \int \md^3k \md \tau \, \frac{z^2}{2} \left(  \tilde \xi_k^{\prime2} - c_s^2 k^2 \tilde \xi_k^2  \right),
\end{align}
where 
\begin{align}
    z^2 = &  2 a^2 \alpha\, \left( \frac{k^2+ 3 \alpha \mathcal{H}^2}{k^2+ \alpha (3-\sigma) \mathcal{H}^2} \right) \label{eq:zsquared_Cuscuton} \\ 
    c_s^2 =& \frac{k^4 + B_1 \mathcal{H}^2 k^2 + B_2 \mathcal{H}^4}{k^4 + A_1 \mathcal{H}^2k^2 + A_2 \mathcal{H}^4} \label{eq:cssquared_Cuscuton}
\end{align}
and 
\begin{align}
    A_1=& 6 \alpha - \alpha \sigma,  \\
    A_2 =& 9 \alpha^2 - 3 \alpha^2 \sigma, \\
    B_1 =& A_1 + \sigma (6 + \eta + \beta - 2\epsilon) + \alpha (\eta - \beta), \\
    B_2 =& A_2 + \alpha \sigma ( 12- 4 \sigma + 3 \eta) + 3 \alpha^2 (\eta-\beta).
\end{align}
First, note that $z^2$ and $c_s^2$ are highly non-local due to the inverse laplacian operators. Second, in the case of $\sigma=0$ $\rightarrow$ $\alpha=\epsilon$, $\eta=\beta$ we recover the standard expression for a single scalar field. 

\begin{figure}[ht]
    \centering
    \includegraphics[scale=0.5]{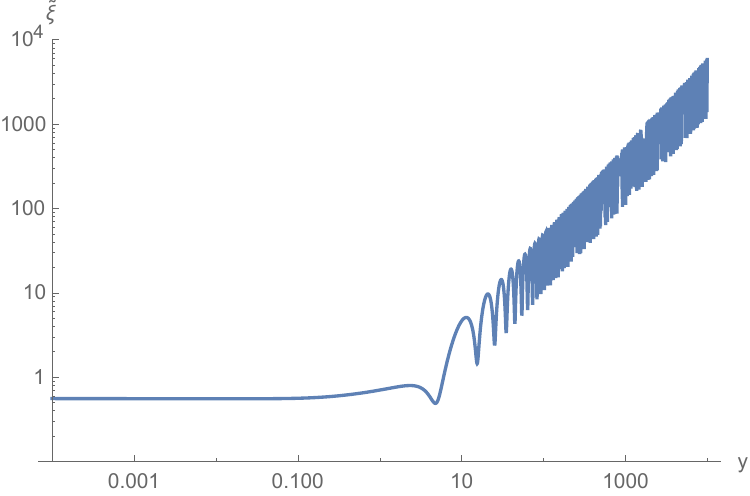}
    \caption{Evolution of the curvature perturbation for the cuscuton in case of power-law with $\epsilon=0.04$ and $\sigma=0.02$. Horizon crossing is at $y=1$.}
    \label{fig:Curvature_conservation}
\end{figure}

To calculate the scalar power spectrum, let us define as usual the canonical normalized Mukhanov-Sasaki variable 
\begin{align}
    v_k = z \tilde \xi_k
\end{align}
in which case the EOM is given by
\begin{align}
v_k^{\prime\prime} + \left(c_s^2 k^2 - \frac{z^{\prime\prime}}{z} \right) v_k =0.
\end{align}
Due to the non-local structure a general solution is quite involved. Instead, let us consider first two different limits. First, in the ultraviolet regime $k/aH\rightarrow \infty$ we recover the standard solution of an harmonic oscillator
\begin{align}
    v_k^{\prime\prime} + k^2 v_k \simeq 0.
\end{align}
On the other hand, in the infrared limit $k/aH \rightarrow 0$
\begin{align}
    \frac{\md}{\md \tau} \left( z^2 \tilde \xi^\prime \right) \simeq 0
\end{align}
which leads for $\epsilon <3$ to a constant and a decaying mode \cite{Boruah:2017tvg}. Therefore, the comoving curvature perturbation is, as usual, conserved on super-horizon scales (see fig. \ref{fig:Curvature_conservation}).

Let us now go into more details by expanding the EOM in the super- and subhorizon limit and then match them both together using that the curvature perturbation is conserved in the infrared limit. 

On subhorizon scales $k^2/(aH)^2 \gg \mathcal{O}(\alpha)$ we expand $z^{\prime\prime}/z$ and $c_s^2$ in orders of the slow-roll parameters. We will assume that the slow-roll parameters are roughly of the same order and will denote it generically as $\mathcal{O}(\alpha)$. Up-to-next-to-leading order we obtain
\begin{align}
   \frac{z^{\prime\prime}}{z} \simeq & \mathcal{H}^2 \left( 2 - \epsilon + \frac{3}{2} \beta \right), \\
   c_s^2 \simeq & 1 + 6 \sigma \frac{\mathcal{H}^2}{k^2}.
   \label{eq:expansion_sound_speed_cuscuton}
\end{align}
The impact of the cuscuton field leads to a modification of the mass term which is, however, slow-roll suppressed.  
Therefore, on subhorizon scales the solutions of the modes up-to-next-to-leading order are given by
\begin{align}
    v_k^{\mathrm{sub}} = \frac{\sqrt{\pi}}{2}\sqrt{y} \left( C_1  H_{\nu_\mathrm{sub}}^{(1)} (\frac{y}{\sqrt{1-2\epsilon}}) + C_2 H_{\nu_\mathrm{sub}}^{(2)}(\frac{y}{\sqrt{1-2\epsilon}})  \right)\, ,
\end{align}
 where $H_{\nu_\mathrm{sub}}^{(1)}$ and $H_{\nu_\mathrm{sub}}^{(2)}$ are the Hankel functions of first and second kind, $y=k/aH$ and
\begin{align}
    \nu_\mathrm{sub} = \frac{\sqrt{9+6\beta - 6 \epsilon - 24 \sigma}}{2 \sqrt{1-2\epsilon}} \simeq \frac{3}{2} + \epsilon + \frac{1}{2} \beta - 2\sigma.
\end{align}
Choosing the Davis-Bunch vacuum deep in the horizon $y\gg 1$ we can fix the initial conditions, $C_1= ((1-2\epsilon) k^2)^{-1/4} e^{i (\nu +1/2) \pi/2 }$, $C_2=0$.

On the other hand, on large superhorizon scales $k^2/(aH)^2 \ll \mathcal{O}(\alpha)$ the EOM is given by 
\begin{align}
    (1-2\epsilon) \frac{\md^2 v^{\mathrm{super}}_k }{\md y^2} - \left( \frac{2 - \epsilon + \frac{3}{2} \beta }{y^2} + \frac{9 \epsilon + 2 \sigma }{9 \alpha} \right) v^{\mathrm{super}}_k \simeq 0
\end{align}
which leads to 
\begin{align}
     v_k^{\mathrm{super}} = \frac{\sqrt{\pi}}{2} \sqrt{y} \left( \tilde c_1 H_{\nu_\mathrm{super}}^{(1)} \left( y \sqrt{\frac{9 \epsilon + 2\sigma}{9 \alpha }}\right) + \tilde c_2 H_{\nu_\mathrm{super}}^{(2)}\left( y \sqrt{\frac{9 \epsilon + 2\sigma}{9 \alpha }}\right) \right)
\end{align}
where 
\begin{align}
    \nu_\mathrm{super} \simeq \frac{3}{2} +\epsilon + \frac{1}{2} \beta.
\end{align}
To obtain $\tilde c_1$ and $\tilde c_2$ we can match the both solutions for the curvature perturbation
\begin{align}
    \tilde \xi_\mathrm{super} ( y_{\star}) = \tilde \xi_\mathrm{sub}(y_\star)
\end{align}
leading to $\tilde c_2=0$ and $\vert \tilde c_1 \vert \simeq  1/\sqrt{k}$.
Since the curvature perturbation is conserved in the infrared limit, the error coming from the matching is slow-roll suppressed. Indeed changing the matching scale from $y_\star=1$ to $y_\star=\alpha$ leads to a deviation of order $\mathcal{O}(\alpha \log \mathcal{\alpha})$. 

Finally, the power spectrum on superhorizon scales can be expressed as
\begin{align}
    P_{\tilde\xi}  = \frac{k^3}{2\pi^2} \vert \tilde \xi_\mathrm{super} \vert^2  \simeq \frac{H^2}{8 \pi^2 \alpha} 
\end{align}
with the spectral index
\begin{align}
    n_s -1 = 3 - 2\nu \simeq -2 \epsilon  - \beta.
    \label{eq:spectral_index_cuscuton}
\end{align}
As a consistency check we can see that for $\sigma \rightarrow 0$ we recover the result from standard single scalar field inflation. Our result is in agreement with the discussion in \cite{Ito:2019fie} focusing on power-law solutions ($\beta=0$).

On the other hand, since the Cuscuton field does not impact the tensor sector at linear order, the power spectrum of the graviational waves is given by the standard form
\begin{align}
    P_\gamma = \frac{2 H^2}{\pi^2}\, ,
\end{align}
with the spectral index
\begin{align}
    n_t = -2 \epsilon
\end{align}
and a tensor-to-scalar ratio 
\begin{align}
    r = 16 \alpha.
    \label{eq:tensor_scalar_ratio_cuscuton}
\end{align}
We can note that the standard consistency relation between the tensor-to-scalar ratio and the spectral index of the tensor modes $r=-8 n_t$ is broken due to the presence of the additional slow-roll parameter $\sigma=\epsilon-\alpha\neq 0$ \cite{Ito:2019fie}. In general, the extra free parameter $\sigma$ enables a wider parameter range for the physical quantities.


\if{}

Due to the non-local structure a general solution is quite involved. Instead, let us consider first two different limits. First, in the ultraviolet regime $k/aH\rightarrow \infty$ we recover the standard solution of an harmonic oscillator
\begin{align}
    v_k^{\prime\prime} + k^2 v_k \simeq 0.
\end{align}
Second, we expand $z^{\prime\prime}/z$ and $c_s^2$ in orders of the slow-roll parameters. We will assume that the slow-roll parameters are roughly of the same order and will denote it generically as $\mathcal{O}(\alpha)$. Up-to-next-to-leading order we obtain
\begin{align}
   \frac{z^{\prime\prime}}{z} \simeq & \mathcal{H}^2 \left( 2 - \epsilon + \frac{3}{2} \beta \right), \\
   c_s^2 \simeq & 1 + 6 \sigma \frac{\mathcal{H}^2}{k^2}.
   \label{eq:expansion_sound_speed_cuscuton}
\end{align}
Note, that we have to assume $ \mathcal{H}^2/k^2  \cdot \mathcal{O}(\alpha) \ll 1$ in order to have a consistent slow-roll expansion, i.e. the expansion breaks down on large super-horizon scales $k^2/(aH)^2 \sim \mathcal{O}(\alpha)$.   
The impact of the cuscuton field leads to a modification of the mass term which is, however, slow-roll suppressed. Therefore, using a de-Sitter approximation (zeroth order in the slow-roll parameters) we recover the usual single scalar field result with a scale invariant spectrum. 

Up-to-next-to-leading order the solutions of the modes are given by 
\begin{align}
    v_k = \frac{\sqrt{\pi}}{2}\sqrt{y} \left( C_1  H_\nu^{(1)} (\frac{y}{\sqrt{1-2\epsilon}}) + C_2 H_\nu^{(2)}(\frac{y}{\sqrt{1-2\epsilon}})  \right)\, ,
\end{align}
 where $H_\nu^{(1)}$ and $H_\nu^{(2)}$ are the Hankel functions of first and second kind, $y=k/aH$ and
\begin{align}
    \nu = \frac{\sqrt{9+6\beta - 6 \epsilon - 24 \sigma}}{2 \sqrt{1-2\epsilon}} \simeq \frac{3}{2} + \epsilon + \frac{1}{2} \beta - 2\sigma.
\end{align}
Choosing the Davis-Bunch vacuum deep in the horizon $y\gg 1$ we can fix the initial conditions, $C_1= ((1-2\epsilon) k^2)^{-1/4} e^{i (\nu +1/2) \pi/2 }$, $C_2=0$.
To evaluate the power spectrum we expand the solution around super-horizon scales, $y\ll 1$. To be precise, we have to consider the scales, $\mathcal{O}(\sqrt{\alpha}) \ll y \ll 1$, since otherwise the slow-roll expansion breaks down. Using the asymptotics for $y\ll 1$ we finally obtain for the curvature perturbation up-to-next-to-leading order
\begin{align}
    \vert \tilde \xi_k \vert \simeq \frac{H}{\sqrt{4 \alpha k^3}} \left( \frac{k}{a H} \right)^{3/2 - \nu} \left( 1 + c_1  \epsilon + c_2 \sigma + c_3 \beta \right)
\end{align}
with 
\begin{align}
    c_1 =  1 - \gamma_E - \log 2, \quad c_2 =  -4 + 2 \gamma_E + 2 \log 2, \quad c_3 = - \frac{1}{4} c_2
\end{align}
where $\gamma_E$ is the Euler-Mascheroni constant.

On the other hand, in the infrared limit $k/aH \rightarrow 0$
\begin{align}
    \frac{\md}{\md \tau} \left( z^2 \tilde \xi^\prime \right) \simeq 0
\end{align}
which leads for $\epsilon <3$ to a constant and a decaying mode \cite{Boruah:2017tvg}. Therefore, the comoving curvature perturbation is, as usual, conserved on super-horizon scales and we can fix the constant solution by matching both solutions at horizon crossing $k=aH$.

The power-spectrum can be expressed as
\begin{align}
    P_{\tilde\xi}  = \frac{k^3}{2\pi^2} \vert \tilde \xi_k \vert^2 \big \vert_{k=aH} \simeq \frac{H^2}{8 \pi^2 \alpha} ( 1 + 2c_1 \epsilon + 2 c_2 \sigma + 2 c_3 \beta)
\end{align}
with the spectral index
\begin{align}
    n_s -1 = 3 - 2\nu \simeq -2 \epsilon + 4 \sigma - \beta.
    \label{eq:spectral_index_cuscuton}
\end{align}
As a consistency check we can see that for $\sigma \rightarrow 0$ we recover the result from standard single scalar field inflation. Note, that the expression for the spectral index differs from the one derived in \cite{Ito:2019fie} in the case of power-law solutions ($\beta=0$) since the correction to the mass term coming from the expansion of the sound speed \eqref{eq:expansion_sound_speed_cuscuton} has not been considered.

On the other hand, since the Cuscuton field does not impact the tensor sector at linear order, the power spectrum of the graviational waves is given by the standard form
\begin{align}
    P_\gamma = \frac{2 H^2}{\pi^2}\, ,
\end{align}
with the spectral index
\begin{align}
    n_t = -2 \epsilon
\end{align}
and a tensor-to-scalar ratio 
\begin{align}
    r = 16 \alpha.
    \label{eq:tensor_scalar_ratio_cuscuton}
\end{align}
We can note that the standard consistency relation between the tensor-to-scalar ratio and the spectral index of the tensor modes $r=-8 n_t$ is broken due to the presence of the additional slow-roll parameter $\sigma=\epsilon-\alpha\neq 0$. In general, the extra free parameter $\sigma$ enables a wider parameter range for the physical quantities.

In \cite{Ito:2019fie} the authors discussed that using the new free parameter the experimental constraints from Planck \cite{Planck:2018jri}
\begin{align}
    n_s \sim 0.96, \qquad r < 0.1.
    \label{eq:Planck_constraints}
\end{align}
can be reconciled with power-law inflation by independently tuning $\alpha$ and $\epsilon$. However, as mentioned before, the corrections to the mass term \eqref{eq:expansion_sound_speed_cuscuton} has not been considered. Including this effect and using the relations for the spectral index \eqref{eq:spectral_index_cuscuton} and the tensor-to-scalar ratio \eqref{eq:tensor_scalar_ratio_cuscuton} we can note that due to the condition $\sigma=\epsilon-\alpha >0$ \eqref{eq:sigma} the power-law model is still inconsistent with the experimental constraints. As we will discuss in the second part, we need to consider generalized cuscuton theories to obtain viable models of power-law inflation. 
\fi

\subsection{Bispectrum}
Using the background EOMs the cubic order of the contribution coming from GR and the cuscuton is given by
\begin{align}
    \delta_3 S_{GR+\varphi} = \int\md^3x & \md t\,a^3 \Big[  3 H^2 \delta N^3 - 9 H^2 \delta N^2 \xi - 6 H \dot \xi \delta N^2 + 18 H \dot \xi \xi  \delta N + 3 \dot \xi^2 ( \delta N - 3 \xi) \nonumber \\
    &+ 2 \frac{\partial^2 \psi}{a^2} (H \delta N - \dot \xi) \left( \delta N -\xi \right) + 2 \frac{\partial_i \xi \partial_i \psi}{a^2} \left( \dot \xi - H \delta N \right) + \delta N \left( - 2 \frac{\partial^2 \xi}{a^2} \xi - \frac{(\partial_i \xi)^2}{a^2} \right)  \nonumber \\
    & + \frac{(\partial_i \xi)^2}{a^2} \xi - 2 \frac{\partial^2 \psi}{a^2} \frac{\partial_i \psi \partial_i \xi}{a^2} + \frac{1}{2} (\delta N - 3\xi) \left( \frac{(\partial^2 \psi)^2}{a^4} - \frac{(\partial_i \partial_j \psi)^2}{a^4}\right)\Big].
\end{align}
The contribution from the inflaton is given by
\begin{align}
    \delta_3 S_\chi = \int \md^3x&\md t\,a^3 \Big[ \frac{1}{2} \dot \chi^2 ( 3\xi \delta N^2- \delta N^3) + \dot \chi \delta \dot \chi ( \delta N^2 - 3 \xi \delta N + \frac{9}{2} \xi^2) - \dot \chi \frac{\partial_i \delta \chi \partial_i \psi}{a^2} (\xi - \delta N) \nonumber \\
    & + \frac{1}{2} \delta \dot \chi^2 ( 3\xi -\delta N) - \delta \dot \chi 
    \frac{\partial_i \delta \chi \partial_i \psi}{a^2} - \frac{1}{2} \frac{(\partial_i \delta \chi)^2}{a^2} ( \delta N +\xi) - U_\chi \delta \chi ( \frac{9}{2} \xi^2 + 3 \xi \delta N) \nonumber \\
    & -  \frac{1}{2} U_{\chi\chi}  \delta \chi^2 ( \delta N + 3 \xi) - \frac{1}{6} U_{\chi\chi\chi} \delta \chi^3\Big].
\end{align}
Note, that we do not need to solve the perturbations of the inflaton up to second order since it multiplies the constraint equation at linear order similar to the second order of the lapse or shift vector.

\subsubsection{Slow-roll approximation}
\label{sec:slow_roll_approximation}
For simplicity, we focus on the expansion of the action in terms of the slow-roll parameters. Thereby, we consider the subhorizon limit $k/(aH) \gg \mathcal{O}(\alpha)$. Since the curvature perturbation is conserved outside the horizon the main contribution to the time integration of the interaction Hamiltonian is coming from the regime around horizon crossing \cite{Maldacena:2002vr} where the approximation is valid. Therefore, this expansion should provide a valid estimate for the bispectrum. By using the solution for $\delta\chi$, $\delta N$ and $\psi$ we can note that $\delta N$, $\psi$ and $\xi$ are of linear order in the slow-roll parameters, namely
\begin{align}
    \xi =& \tilde \xi + \frac{H}{\dot\chi} \delta \chi \simeq - 3 H^2 \alpha a^2 \partial^{-2} \tilde\xi + \alpha a^2 H \partial^{-2} \dot{\tilde\xi}, \\
    \delta N =& \frac{\dot \xi}{H} + \frac{\dot\chi}{2H} \delta \chi \simeq - 6 H^2 \alpha a^2 \partial^{-2} \tilde \xi - \alpha H a^2 \partial^{-2} \dot{\tilde \xi} + \alpha a^2 \partial^{-2} \ddot{\tilde\xi} - \alpha \tilde\xi, \\
    \psi =& - \frac{\xi}{H} + \frac{a^2}{\partial^2} \left( \alpha \dot \xi - \frac{\dot\chi}{2H} \delta \dot \chi + \frac{1}{2} \dot\chi  \left( \frac{1}{2} \beta - \sigma \right)  \delta \chi \right) \simeq 3 H \alpha a^2 \partial^{-2} \tilde \xi.
\end{align}
Therefore, it is immediate to see that the contributions to the cubic action from GR and the cuscuton are at least of cubic order in the slow-roll parameters and cannot contribute to the leading contribution at quadratic order. 

Since $\delta \chi$ scales as $\mathcal{O}(\sqrt{\alpha})$ (see eq.~(\ref{eq:solution_chi_linear})) the leading contributions from the inflaton at quadratic order in the slow-roll parameters are given by
\begin{align}
     \delta_3 S_\chi\simeq \int \md^3x&\md t\,a^3 \Big[ \frac{1}{2} \delta \dot \chi^2 ( 3\xi -\delta N) - \delta \dot \chi 
    \frac{\partial_i \delta \chi \partial_i \psi}{a^2} - \frac{1}{2} \frac{(\partial_i \delta \chi)^2}{a^2} ( \delta N +\xi) \Big].
\end{align}
Inserting the solutions for the shift and lapse function and for the inflaton we obtain up to quadratic order in the slow-roll parameters 
\begin{align}
    \delta_3 S \simeq \int \md^4x\,a^3 \alpha^2 \Big[ \dot{\tilde\xi}^2 \tilde \xi + \tilde \xi \frac{(\partial_i \tilde \xi)^2}{a^2} - 2 \dot{\tilde\xi} \partial^k\tilde\xi \partial_k \partial^{-2} \dot{\tilde\xi} + \frac{\xi}{\alpha H} \dot{\tilde\xi} \frac{\delta L}{\delta \tilde\xi} \Big],
\end{align}
where $\delta L/\delta \tilde\xi$ is the EOM of the curvature perturbation of linear order at leading order in the slow-roll parameters
\begin{align}
    \frac{\delta L}{\delta \tilde\xi} = \ddot{\tilde\xi} + 3 H \dot{\tilde\xi} - \frac{\partial^2 \tilde\xi}{a^2}.
\end{align}
Note, that the previous expression coincides with the standard one from a single scalar field inflation except for the prefactor in front of $\delta L/\delta \tilde \xi$. Note, that for power-law solutions where $\eta=0$ the result for the bispectrum is the same as the one for single-field slow-roll models of inflation up to leading order in the slow-roll parameters. To conclude, as it is the case for the canonical single-field models the non-gaussianties are suppressed by the slow-roll parametersand, therefore, outside the reach of the current experimental constraints from {\it Planck} \cite{Planck:2018jri}.

\bigskip

\noindent {\textbf{Breakdown of the slow-roll expansion at higher order} }
\medskip

\noindent
Before, we discuss the generalized cuscuton models let us discuss potential issues in the calculation of the bispectrum at higher order in slow-roll due to the slow-roll expansion. In general, due to the non-local expansion of the inflaton field 
\begin{align}
    \delta \chi \simeq - \frac{\dot \chi}{H} \tilde \xi + \frac{\dot \chi}{H} \sum_n \left( b_n \left( H^2\frac{a^2}{\partial^2} \right)^n \tilde \xi + d_n \left( H^2\frac{a^2}{\partial^2} \right)^n \dot{\tilde \xi}  \right)
\end{align}
where $b_n$ and $d_n$ are the expansion coefficients which are of the order $\mathcal{O}(\alpha^n)$, the cubic action at higher order in the slow-roll expansion will contain terms with more and more inverse laplacian operators. Indeed, at cubic order in the slow-roll parameters there are already interaction terms with a structure like
\begin{align}
   \int  \md^4 x  a^7 \alpha^3 H^5 \dot{\tilde\xi} (\partial^{-2} \tilde\xi)^2, \quad \int  \md^4 x  a^7 \alpha^3 H^6 {\tilde\xi} (\partial^{-2} \tilde\xi)^2.
\end{align}
Computing the three point function terms like that will diverge in the asymptotic limit $\tau \rightarrow 0$ since
\begin{align}
     \lim_{\tau \rightarrow 0} \left(\frac{a^2 H^2} {k^2}\right)^n \sim  \lim_{\tau \rightarrow 0} (\tau^2 k^2)^{-n} \rightarrow \infty.
\end{align}
This is, however, exactly, the regime where our initial slow-roll expansion breaks down $\lim_{\tau \rightarrow 0} k/aH \rightarrow 0$.
Therefore, this is not a physical problem but instead an artefact of our expansion. At higher order in slow-roll parameters we need to be more careful in choosing the correct normalization for the integration limits. This problem is not unique for cuscuton inflation but similar issues occur in generic non-local theories of inflation (see, e.g. \cite{Barnaby:2008fk,Seery:2008qj}). 

\subsubsection{Scalar-tensor interactions}
For completeness reason let us also shortly discuss the impact of the cuscuton field on the tensor-scalar interactions. Note, that the tensor-tensor-tensor three point function will be the same as in single scalar field inflation since the cuscuton field does not directly impact the tensor modes at this order.  

Therefore, let us first consider the tensor-scalar-scalar coupling. At leading order in the slow-roll parameters there is only one term which contributes, namely
\begin{align}
    S_{\gamma\tilde\xi\tilde\xi} \simeq \int \md^4x\, a \frac{1}{2} \gamma^{ij} \partial_i \delta \chi \partial_j \delta \chi.
\end{align}
Using the solution for $\delta \chi$ we obtain
\begin{align}
    S_{\gamma\tilde \xi \tilde\xi} \simeq \int \md^4x\, a \alpha \gamma^{ij} \partial_i \tilde \xi \partial_j \tilde \xi
\end{align}
which agrees with the results from standard single scalar field inflation. Similarly, we obtain for the scalar-tensor-tensor coupling
\begin{align}
    S_{\gamma\gamma\tilde\xi} =& \int \md^4x\, a^3 \Big[ \frac{1}{8} \dot \gamma_{ij}^2 ( 3 \xi -\delta N) - \frac{1}{4} \dot \gamma_{ij} \partial_k \gamma^{ij} \frac{\partial^k \psi}{a^2}- (\delta N +  \xi ) \frac{1}{8} \frac{(\partial_k \gamma_{ij})^2}{a^2} \Big] \nonumber \\
    \simeq & \int \md^4x\, a^3 \Big[ \frac{\alpha}{8}  \left(\dot \gamma_{ij}^2 + \frac{(\partial_k \gamma_{ij})^2}{a^2} \right) - \frac{\alpha}{4} \dot \gamma_{ij} \partial_k \gamma^{ij} \partial^k \partial^{-2} \dot{\tilde\xi} + \frac{1}{4 H} \xi \dot \gamma_{ij} \frac{\delta L}{\delta \gamma_{ij}} \Big] 
\end{align}
where $\delta L/\delta \gamma_{ij}$ is the linear equation of motion for the tensor perturbations. 
\bigskip

To summarize the first part, we can conclude that the presence of the cuscuton field does not spoil the inflationary paradigm of single-scalar field inflation but instead the impact of the non-dynamical cuscuton field is slow-roll suppressed which makes it challenging to test. Further, possible signatures as the modification of the scalar spectral index are, in general, degenerate. Therefore, we would need a direct detection of the primordial gravitational wave power spectrum in order to disentangle it and to confirm the need for a non-trivial inflationary model.

\section{Generalized Cuscuton}
\label{sec:Generalized_Cuscuton}
In \cite{Gao:2019twq} the authors constructed a spatial covariant gravity theory up to the quadratic level in derivatives of the metric with just two tensor degrees of freedom which can be understood as a generalization of the cuscuton model in the unitary gauge. The action is given by
\begin{align}
    S = \frac{1}{2} \int \md^4x\,\sqrt{h} N \Big[ & \frac{N}{b_2+N} K_{ij} K^{ij} - \frac{1}{3} \left( \frac{2N}{b_1+N} + \frac{N}{b_2+N} \right) K^2 + \rho_1 + (1 + \rho_2) \bar R  \nonumber \\
    & + \frac{1}{N} (\rho_3 + \rho_4 \bar R) \Big]\, ,
\end{align}
where $b_2$, $b_1$, $\rho_i$ are free functions of time. We can recover the standard cuscuton model by setting $\rho_2=b_2=b_1=\rho_4=0$, $\rho_1=-2V(\varphi(t))$ and $\rho_3 = 2 \mu^2 \vert \dot \varphi(t) \vert$. For $b_1 \neq b_2$ the action is outside the beyond Horndeski (GLPV) class \cite{Gleyzes:2014dya}. The ansatz can be further generalized by including terms linear in the extrinsic curvature. For simplicity, we do not discuss this possibility here.  Note that the action slightly differs from the commonly discussed extended cuscuton models \cite{Iyonaga:2018vnu,Quintin:2019orx,Iyonaga:2020bmm} which corresponds to the case $b_1=b_2$. In \cite{Iyonaga:2021yfv} the model has been discussed in the context of black hole solutions and the late universe. 

Adding the minimally coupled scalar field to describe the inflaton, the background equations of motion are given by
\begin{align}
    \frac{3 H^2}{(1+b_1)^2} =& \frac{1}{2} \dot\chi^2 + U(\chi) - \frac{1}{2} \rho_1, \\
    - 2 \dot H =& (1+b_1)^2 \left(\frac{1}{2} \dot \chi^2 + U - \frac{1}{2} \rho_1\right) + (1+b_1) \left( \frac{1}{2} (\rho_1+\rho_3 + \dot \chi^2) -U \right) - \frac{2 H}{1+b_1} \dot b_1.
\end{align}
Note, that the parameters $\rho_2$, $\rho_4$ and $b_2$ do not impact the background evolution. While in the original cuscuton model the tensor modes are not impacted at linear level this is not anymore the case but instead
\begin{align}
    \delta_2 S = \frac{1}{8}\int \md^3x\,\md t\, a^3 \frac{1}{1+b_2} \left( \dot\gamma_{ij}^2 - (1 + \rho_2+\rho_4)(1+b_2) \frac{(\partial_k \gamma_{ij})^2}{a^2} \right).
\end{align}
In order to avoid ghost or gradient instabilities we have to require that
$1+b_2 >0$ and $1+\rho_2+\rho_4>0$.

Similarly, the linear scalar perturbations can be expressed as
\begin{align}
    \delta_2 S = & \int \md^3x \md t\,  a^3 \Big[ (1+\rho_2 + \rho_4) \frac{(\partial_k \xi)^2}{a^2} - \left(\frac{3}{(1+b_1)^3} H^2 - \frac{1}{2} \dot \chi^2\right) \delta N^2 - \frac{3}{(1+b_1)} \dot \xi^2 \nonumber \\ &+ \frac{6 H}{(1+b_1)^2} \delta N \dot \xi  + \frac{2}{b_1+1} \dot \xi \frac{\partial^2 \psi}{a^2} - \frac{2 H}{(1+b_1)^2 } \delta N \frac{\partial^2 \psi}{a^2} + \frac{1}{3} \left( \frac{1}{1+b_2} - \frac{1}{1+b_1}\right) \frac{(\partial^2 \psi)^2}{a^4} \nonumber \\
    & - 2 (1+\rho_2) \delta N \frac{\partial^2\xi}{a^2}  + \frac{1}{2} \delta \dot\chi^2 - \frac{1}{2} \frac{(\partial_k \delta\chi)^2}{a^2} + \dot \chi \delta \chi \frac{\partial^2 \psi}{a^2} - \dot \chi \delta N \delta\dot \chi + 3 \dot \chi \xi \delta\dot \chi \nonumber \\
    & - U_\chi \delta \chi (\delta N + 3 \xi) - \frac{1}{2} U_{\chi\chi}\delta\chi^2 \Big].
    \label{eq:Linear_Perturbations_Generalized_Cuscuton}
\end{align}
As mentioned before, for $b_1 \neq b_2$ the action is outside the beyond Horndeski (GLPV) class which can lead to higher order terms in spatial derivatives in the linear perturbations due to the term proportional to $(\partial^2\psi)^2$ (see for instance \cite{Langlois:2017mxy} where the kinetic detuning $b_1\neq b_2$ corresponds to the parameter $\alpha_L \neq0$).  In the following, we will split our discussion into the two different cases
\begin{itemize}
    \item inside the GLPV class:  $b_2=b_1$
    \item outside the GLPV class: $b_2 \neq b_1 $ 
\end{itemize}
A full analysis with generic free functions $b_i(t)$ and $\rho_j(t)$ is beyond the scope of the paper. Instead, we will analyze the case where the background evolution is identical to the original cuscuton model, $b_1=0$, $\rho_1=-2V(\varphi(t))$ and $\rho_3 = 2 \mu^2 \vert \dot \varphi(t) \vert$ so that we can use the power-law solutions constructed in \cite{Ito:2019fie}. Further, we will assume that $\rho_2$, $\rho_4$ and $b\equiv b_2$ are constant in time.

\subsection{Linear perturbations and power spectrum}
\label{sec:Linear_perturbations_Generalized_Cuscuton}

Using the EOMs for the non-dynamical variables and redefining the curvature perturbation \eqref{eq:Redefining_curvature} to integrate out the inflaton field (see the appendix \ref{app:Linear_perturbations_generalized_Cuscuton} for the details) we can write the second order action of the scalar perturbations as
\begin{align}
    \delta_2 S= \int \md^3k \md \tau\,\frac{1}{2} z^2(k,\tau) \Big[ \tilde\xi^{\prime 2} - c_s^2(k,\tau) \frac{k^2}{a^2} \tilde \xi^2 \Big]\, ,
\end{align}
where 
\begin{align}
    z^2(k,\tau) =& 2 a^2 \alpha \frac{  b (1+\rho_2)^2 k^4 + d_1 \mathcal{H}^2 k^2 + d_2 \mathcal{H}^4 }{ b (1+\rho_2)^2 k^4 + d_3 \mathcal{H}^2 k^2 + d_4 \mathcal{H}^4 }, \\
    c_s^2(k,\tau) =& \frac{b^2 (1+\rho_2)^4 k^8 + b d_5 \mathcal{H}^2 k^6 + d_6 \mathcal{H}^4 k^4 + d_7 \mathcal{H}^6 k^2 + d_8 \mathcal{H}^8}{b^2(1+\rho_2)^2 k^8 + b d_9 \mathcal{H}^2 k^6 + d_{10} \mathcal{H}^4 k^4 + d_{11} \mathcal{H}^6 k^2 + d_{12}  \mathcal{H}^8}\, ,
\end{align}
and the explicit form of $d_i$ are given in the appendix.
Using the Mukhanov-Sasaki variables $v_k=z\tilde \xi$ the EOM are then as usual given by 
\begin{align}
     v_k^{\prime\prime} + \left(c_s^2 k^2 - \frac{z^{\prime\prime}}{z} \right) v_k=0.
\end{align}

\subsubsection*{Inside GLPV}
Inside the GLPV class, $b=0$, the non-local structure of $z^2$ and $c_s^2$ simplifies substantially. In fact the fundamental form is now the same as in the original cuscuton model (see \eqref{eq:zsquared_Cuscuton} and \eqref{eq:cssquared_Cuscuton}). This will be the case as long as we are inside the GLPV class and does not change if we introduce $b_1=b_2\neq 0$ and consider generic time dependent functions.

We can note that for the case $\sigma=0$ we do not recover anymore the standard single scalar field but instead 
\begin{align}
   z^2 =& 2 a^2 \epsilon, \\
   c_s^2=& \frac{\rho_4 + \epsilon \rho_2 (1+\rho_2)}{\rho_4 + \epsilon \rho_2 }.
\end{align}
Even if there are no modification at the background the impact of the generalized cuscuton will reappear at linear and higher order. However, in this case the non-local structure vanishes and it is straightforward to calculate the power spectrum. 

Let us now go back to the cuscuton background $\sigma\neq 0$. We can note that in the ultraviolet limit $k/aH\rightarrow \infty$, the sound speed
\begin{align}
    c_s^2 \simeq \frac{\alpha (1+\rho_2)^2 - \epsilon (1+\rho_2) + \rho_4 }{\alpha (1+2\rho_2) - \epsilon (1+\rho_2) +\rho_4} 
\end{align}
will, in general, not anymore be equivalent to the speed of light. Similar to the cuscuton case we can expand the mass and the sound speed in orders of the slow-roll parameters in the sub- and superhorizon limit and match both solutions together.

In the subhorizon limit $k/(aH)\gg \mathcal{O}(\alpha)$ we obtain
\begin{align}
    \frac{z^{\prime\prime}}{z} \simeq & (-2+\epsilon) \mathcal{H}^2, \\
    c_s^2 \simeq & \tilde c_s^2 + \tilde m^2 \frac{\mathcal{H}}{k^2}
\end{align}
where $\tilde m = \mathcal{O}(\alpha)$ and 
\begin{align}
    \tilde c_s^2 = \begin{cases}
     1 + \alpha \frac{\rho_2^2}{\rho_4} & \text{if $\rho_4\neq 0$} \\
     \frac{(1+\rho_2) (-\sigma + \alpha \rho_2)}{\alpha (1+2 \rho_2) -\epsilon (1+\rho_2)} & \text{if $\rho_4=0$}
    \end{cases}.
\end{align}
While in the general case $\rho_4\neq0$ the correction to the sound speed are slow-roll suppressed in the specific case $\rho_4=0$ the sound speed can deviate substantially from the speed of light. Note, that since $\tilde m = \mathcal{O}(\alpha)$ the modified mass does not impact the matching procedure up-to-leading order.

On the other hand, in the superhorizon limit $k/(aH) \ll \mathcal{O}(\alpha)$
\begin{align}
   (1-2\epsilon) \frac{\md v_k^{\mathrm{super}}}{\md y^2} + \left(  \frac{-2 +\epsilon}{y^2} + y^2 \left( - \frac{11 \rho_4}{9 \alpha} + \frac{-2 \alpha + 11 (1+\rho_2) \epsilon - \rho_2 \alpha }{9 \alpha} \right) \right) v^{\mathrm{super}}_k =0.
\end{align}
Note, that we have to assume that $\rho_4 < 0$ in order to avoid a singularity in $z^2$. Further, if $\rho_4\neq0$ the term proportional to $y^2$ is of order $\mathcal{O}(1/\alpha)$. However, following the matching procedure in section  \ref{sec:Linear_perturbations_Cuscuton} the correction is still suppressed by $\mathcal{O}(\alpha \log \alpha)$.
The power spectrum at superhorizon scales is given by
\begin{align}
    P_{\tilde \xi} = \frac{H^2}{8 \pi^2 \alpha \tilde c_s}
\end{align}
where
\begin{align}
    n_s= -2\epsilon.
\end{align}
In the general case $\rho_4 \neq 0$ the sound speed $\tilde c_s \simeq 1$ and, consequently, there is no difference in comparison to standard cuscuton up-to-leading order. 

On the other hand, the power spectrum of the gravitational waves gets impacted by the modification of the propagation speed
\begin{align}
    P_\gamma  \simeq \frac{2 H^2}{\pi^2 c_t } = \frac{2 H^2}{\pi^2 \sqrt{1+\rho_2 + \rho_4}}\, ,
\end{align}
so that the tensor-to-scalar perturbation ratio is given by
\begin{align}
    r = 16 \alpha \frac{ \tilde c_s}{c_t}.
\end{align}

\if{}
Inside the GLPV class, $b=0$, the non-local structure of $z^2$ and $c_s^2$ simplifies substantially. In fact the fundamental form is now the same as in the original cuscuton model (see \eqref{eq:zsquared_Cuscuton} and \eqref{eq:cssquared_Cuscuton}). This will be the case as long as we are inside the GLPV class and does not change if we introduce $b_1=b_2\neq 0$ and consider generic time dependent functions.

However in the ultraviolet limit $k/aH\rightarrow \infty$, the sound speed
\begin{align}
    c_s^2 \simeq \frac{\alpha (1+\rho_2)^2 - \epsilon (1+\rho_2) + \rho_4 }{\alpha (1+2\rho_2) - \epsilon (1+\rho_2) +\rho_4} 
\end{align}
will, in general, not anymore be equivalent to the speed of light.
Similarly to the discussion in the cuscuton model, we can analyze the slow-roll approximation by expanding $c_s^2$ and the mass term $z^{\prime\prime}/z$ in orders of the slow-roll parameters assuming that $k^2/(aH)^2 \gg \mathcal{O}(\alpha)$ and $\rho_2 \sim \rho_4 \sim \mathcal{O}(1)$, which leads to
\begin{align}
    \frac{z^{\prime\prime}}{z} \simeq &  \mathcal{H}^2 (2-\epsilon), \\
    c_s^2 \simeq & 1 + \frac{\alpha \rho_2^2}{\rho_4} + 6 \sigma \frac{\mathcal{H}^2}{k^2}. \label{eq:sound_speed_inside_GLPV}
\end{align}
First, we can note that at zeroth order in the slow-roll parameters (de-Sitter approximation) we recover the single scalar field inflation. At next-to-leading order the impact of the new free parameters $\rho_2$ and $\rho_4$ will lead to a modification of the sound speed while the mass term will remain unchanged w.r.t to cuscuton model (notice that we are considering power-law solutions for which $\eta=\beta=0$). If $\rho_2=0$ there are even no distinctions between the cuscuton model and the generalized version up to the next-to-leading order. There is one specific case, namely, where $\rho_2 \neq 0$ and $\rho_4=0$. In this case the approximation for the sound speed \eqref{eq:sound_speed_inside_GLPV} does not work, instead, we can see that the limit $\alpha \equiv \alpha x $ and  $\epsilon \equiv \epsilon x$ with $x\rightarrow 0$ leads to
\begin{align}
  \lim_{x\rightarrow 0}  c_s^2\vert_{\rho_4=0} = \frac{(1+\rho_2) (\alpha (1+\rho_2) - \epsilon )}{\alpha (1+2 \rho_2) - \epsilon (1+\rho_2)}.
\end{align}
Therefore, we obtain already a modification at zeroth order. In the following we will not consider this case anymore but instead assume that $\rho_2=0$ if $\rho_4=b=0$.

Following the procedure in section \ref{sec:Linear_perturbations_Cuscuton} we can calculate the power spectrum up-to-next-to leading order 
\begin{align}
    P_{\tilde \xi} = \frac{H^2}{8 \pi^2 \alpha \tilde c_s} (1 + 2c_1 \epsilon + 2 c_2 \sigma ) \simeq \frac{H^2}{8 \pi^2 \alpha }  \left(1 + 2 c_1 \epsilon + 2 c_2 \sigma - \frac{1}{2} \alpha \frac{\rho_2^2}{\rho_4}\right) 
\end{align}
where 
\begin{align}
    \tilde c_s^2 = 1 + \alpha \frac{\rho_2^2}{\rho_4}. 
\end{align}
The spectral index does not change in comparison to the cuscuton model
\begin{align}
    n_s -1 = 3 - 2 \nu = -2 \epsilon + 4 \sigma\, ,
\end{align}
taking into account that we are consider power law solutions where $\eta=\beta=0$.
On the other hand, the power spectrum of the gravitational waves gets impacted by the modification of the propagation speed
\begin{align}
    P_\gamma  \simeq \frac{2 H^2}{\pi^2 c_t } = \frac{2 H^2}{\pi^2 \sqrt{1+\rho_2 + \rho_4}}\, ,
\end{align}
so that the tensor-to-scalar perturbation ratio is given by
\begin{align}
    r = 16 \frac{\alpha}{\sqrt{1+\rho_2 + \rho_4}}.
\end{align}
Therefore, in order to fulfill the experimental constraints on the scalar spectral index and the tensor-to-scalar ratio \eqref{eq:Planck_constraints} we have to require superluminal propagation of the gravitational waves $c_t=\sqrt{1+\rho_2+\rho_4}>1$.
\fi

\subsubsection*{Outside GLPV}
Let us now go back to the general case including the kinetic detuning, $b\neq 0$. We can note that in contrast to the cuscuton model the non-local structure of $z^2$ and $c_s^2$ contain higher order polynomials in $k$. However, in the ultraviolet limit $k/aH\rightarrow \infty$ we recover the result from the single scalar field inflation
\begin{align}
    z^2 \simeq & 2 a^2 \alpha, \\
    c_s^2 \simeq & 1.
\end{align}
Contrary to what we might expect the kinetic detuning does not lead to a higher order dispersion relation in the ultraviolet limit.

If $\sigma=0$ at the background we again do not recover standard single scalar field. But in comparison to the previous case the non-local structure of $z^2$ and $c_s^2$ remains.

We have already seen that in the ultraviolet limit we recover the standard result. At subhorizon scales $k/(aH) \ll $ the EOM can be written as
\begin{align}
   (1-2\epsilon) \frac{\md^2 v_k^{\mathrm{super}}}{\md y^2} + \left( \frac{-2 + \epsilon}{y^2} - \frac{11 \rho_4}{9 \alpha} \right) v_k^{\mathrm{super}} \simeq 0
\end{align}
which is the same as for $b=0$. Matching the solutions together the power spectrum on superhorizon scales is, therefore, given by
\begin{align}
    P_{\tilde \xi} = \frac{H^2}{8 \pi^2 \alpha }
\end{align}
with 
\begin{align}
    n_t = -2 \epsilon.
\end{align}
Up-to-leading order the result is equivalent to the standard cuscuton case. 
On the other hand, the tensor power spectrum 
\begin{align}
    P_h\vert_{\rho_4=0}= \frac{2 H^2 (1+b)}{\pi^2 c_t} = \frac{2 H^2}{\pi^2 } \sqrt{\frac{1+b}{1+\rho_2}},
\end{align}
so that the tensor to scalar ration $r$ is given by
\begin{align}
    r = 16 \alpha \sqrt{\frac{1+b}{1+\rho_2}}.    
\end{align}

\begin{figure}
    \centering
    \includegraphics[scale=0.5]{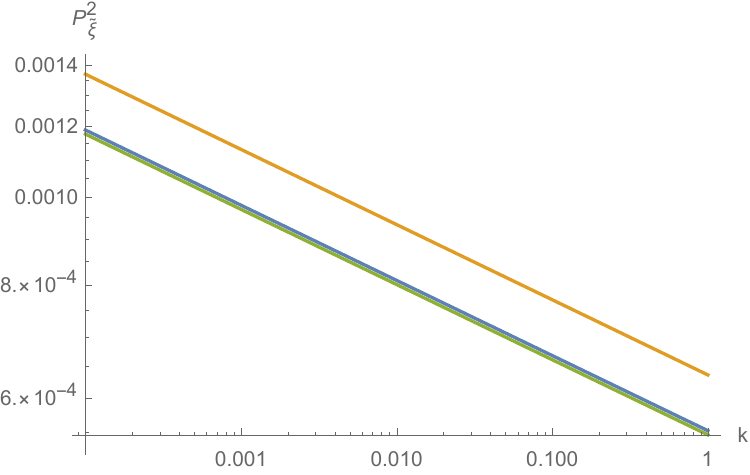}
    \caption{The Power spectrum in generic units for cuscuton (orange), inside GLPV (blue) and outside GLPV (green) for $\epsilon=0.04$, $\sigma=0.02$ $b=1$, $\rho_4=-0.5$ and $\rho_2=0.5$. The difference in the amplitude is of the order $\epsilon \log \epsilon \simeq 0.87$.}
    \label{fig:Power_spectrum}
\end{figure}

Summarizing the impact of the modifications of the cuscuton model on the scalar power spectrum is mostly slow-roll suppressed (see fig. \ref{fig:Power_spectrum}).  There is only one specific subcase inside the GLPV class with a leading order correction coming from the deviation of the sound speed at the ultraviolet limit $\tilde c_s$. Therefore, the tensor power spectrum is the best way to distinguish these different scenarios up-to-leading order. The reason behind it is that the modifications coming from the (generalized) cuscuton are important on scales $k/(aH) \sim \mathcal{O}(\alpha)$ which are, however, not relevant for the power spectrum up-to-leading order. 

\if{}
Let us now go back to the general case including the kinetic detuning, $b\neq 0$. We can note that in contrast to the cuscuton model the non-local structure of $z^2$ and $c_s^2$ contain higher order polynomials in $k$. However, in the ultraviolet limit $k/aH\rightarrow \infty$ we recover the result from the single scalar field inflation
\begin{align}
    z^2 \simeq & 2 a^2 \alpha, \\
    c_s^2 \simeq & 1.
\end{align}
Contrary to what we might expect the kinetic detuning does not lead to a higher order dispersion relation in the ultraviolet limit.

Expanding the mass term and the sound speed in orders of the slow-roll parameters the EOM can be expressed as 
\begin{align}
     v_k^{\prime\prime} + \left( k^2 - \mathcal{H}^2 (2-\epsilon)   + m^2(k,\tau) \mathcal{H}^2  \right) v_k =0
\end{align}
where 
\begin{align}
    m^2(k,\tau) =  \frac{f_1 k^6 + f_2 k^4 \mathcal{H}^2 + f_3 k^2 \mathcal{H}^4 + f_4 \mathcal{H}^6 }{(b (1+\rho_2)^2 k^2 - 3 \rho_4 \mathcal{H}^2)^3} 
\end{align}
with 
\begin{align}
    f_1 =& - b^2 (1+\rho_2)^4 \left( 3 \alpha \rho_2^2 - 6 b \epsilon (1+ \rho_2)^2 + b \alpha (6 + 9 \rho_2 + 3 \rho_2^2+ \rho_4 ) \right), \\
    f_2 = & - b (1+\rho_2)^2 \rho_4 \big( - 18 \alpha \rho_2^2 + 5 b^2 \alpha (1+ \rho_2)^2 + 54 b \epsilon (1+\rho_2)^2 \nonumber \\
    &- 6 b \alpha (9 + 16 \rho_2 + 7 \rho_2^2 + \rho_4) \big), \\
    f_3 = & 3 \rho_4^2 (- 9 \alpha \rho_2^2 + \alpha b^2 (1+\rho_2)^2 + 54 b \epsilon (1+\rho_2)^2 - 3 b \alpha (18 + 35 \rho_2 + 17 \rho_2^2 + \rho_4), \\
    f_4 = & - 162 (\epsilon-\alpha) \rho_4^3.
\end{align}
Interestingly, we can note, that the sound speed is not impacted by the modifications up-to-next-to-leading order.
Since $m^2(k,\tau)\sim \mathcal{O}(\alpha)$ at the de-Sitter approximation we recover the single scalar field. However, at next-to-leading order the structure of the mass term is much more involved. A general analytical solution is beyond the scope of this paper. Let us, instead, discuss some simple special cases. If $\rho_2=\rho_4=0$ we can note that we recover the EOM of the original cuscuton model up-to-the-next-to-leading order. If just $\rho_4=0$ the mass term simplifies to 
\begin{align}
    m^2(k,\tau) \vert_{\rho_4=0} =  -\frac{3 \alpha \rho_2^2+ b(1+\rho_2) (-6 \epsilon (1+\rho_2) + 3 \alpha (2+\rho_2))}{b(1+\rho_2)^2}\, ,
\end{align}
resulting in a scale-independent mass term. Therefore, the power-spectrum up-to-next-to-leading order is given by
\begin{align}
    P_{\tilde\xi} \vert_{\rho_4=0} = \frac{ H^2}{8 \pi^2 \alpha } \left( 1 + 2c_1 \epsilon -  \frac{1}{3} c_2 m^2\vert_{\rho_4=0}  \right)\, ,
\end{align}
and the spectral index by
\begin{align}
    n_s\vert_{\rho_4=0} -1 = 3- 2 \nu = - 2 \epsilon - \frac{2}{3} m^2 \vert_{\rho_4=0}.
\end{align}
The tensor sector is now given by
\begin{align}
    P_h\vert_{\rho_4=0}= \frac{2 H^2 (1+b)}{\pi^2 c_t} = \frac{2 H^2}{\pi^2 } \sqrt{\frac{1+b}{1+\rho_2}},
\end{align}
which leads to the tensor-to-scalar ratio
\begin{align}
    r\vert_{\rho_4=0} = 16 \alpha \sqrt{\frac{1+b}{1+\rho_2}}.
\end{align}
In that case it is possible to fulfill the experimental constraints \eqref{eq:Planck_constraints} without the need of superluminal tensor propagation by appropriately tuning $b$ and $\rho_2$. 
\fi

\subsection{Bispectrum}
Let us discuss the impact of the generalized cuscuton model on the bispectrum. Excluding the subcase $b=\rho_4=0$ and $\rho_2\neq 0$ (see the discussion in section \ref{sec:Linear_perturbations_Generalized_Cuscuton}) \eqref{eq:delta_N_GC}, \eqref{eq:psi_GC} and \eqref{eq:delta_chi_GC} indicate
that similar to the cuscuton model the linear perturbations of the metric are slow-roll suppressed (cubic in the slow-roll parameters).  Therefore, up-to-leading order the only relevant terms are coming from 
\begin{align}
     \delta_3 S_\chi\simeq \int \md^3x&\md t\,a^3 \Big[ \frac{1}{2} \delta \dot \chi^2 ( 3\xi -\delta N) - \delta \dot \chi 
    \frac{\partial_i \delta \chi \partial_i \psi}{a^2} - \frac{1}{2} \frac{(\partial_i \delta \chi)^2}{a^2} ( \delta N +\xi) \Big].
\end{align}
Using 
\begin{align}
     \delta N =& \frac{\dot \xi}{H} + \frac{\dot \chi}{2 H} \delta \chi + \frac{b }{3 (1+b)} \frac{k^2}{a^2} \psi, \\
    \frac{k^2}{a^2} \psi \simeq & - \frac{3 (1+b)}{2 H^2 (6 + 2 b \alpha)} \Big(  -2 H \dot \chi \delta \dot \chi + 4 (1+\rho_2) H \frac{k^2}{a^2} \xi  \Big), \\
    \delta \chi \simeq & - \frac{\dot \chi}{H} \tilde \xi
\end{align}
we can express the action as
\begin{align}
    \delta_3 S \simeq &\int \md^3x\md t\, a^3  \Big[ \alpha^2 \dot{\tilde\xi}^2 \tilde\xi + \alpha^2 \frac{(\partial_i \tilde\xi)^2}{a^2} - 2 \alpha^2 (1+b) \dot{\tilde\xi} \partial_i \tilde\xi \partial_i \partial^{-2} \dot{\tilde\xi} + \frac{\alpha^2 b}{3 H} \dot{\tilde\xi} \left( \dot{\tilde\xi}^2 + \frac{(\partial_i \tilde\xi)^2}{a^2}\right) \nonumber \\
    & - \frac{\alpha b}{3 H^2} (1+ \rho_2) \left( \dot{\tilde\xi}^2 + \frac{(\partial_i \tilde\xi)^2}{a^2} \right) \frac{\partial^2 \xi}{a^2}+ 2 \alpha (\rho_2 + b + b\rho_2) \frac{\dot{\tilde\xi}}{H} \frac{\partial_i \tilde\xi}{a^2} \partial_i \xi \nonumber \\
    & +  2 \alpha \xi \frac{\dot{\tilde\xi}}{H} \frac{\delta \L}{\delta \tilde\xi}\Big].
\end{align}
Since at leading order
\begin{align}
    \xi \simeq \frac{3 \alpha  \rho_2 a^2 H^2 }{b(1+\rho_2)^2 \partial^2 + 3 \rho_4 a^2 H^2} \tilde\xi + \frac{\alpha  b(1+\rho_2) H a^2}{b(1+\rho_2)^2\partial^2 + 3 \rho_4 a^2 H^2} \dot{\tilde\xi} + \mathcal{O}(\alpha^2)\, , 
\end{align}
the terms in the second line are in general non-local operators. 
Let us first focus on the simpler case inside the GLPV class with $b=0$.

\subsubsection*{Inside GLPV}
If $b=0$ the relation for $\xi$ simplifes to
\begin{align}
    \xi \simeq \alpha \frac{\rho_2}{\rho_4} \tilde\xi + \mathcal{O}(\alpha^2)\, ,
\end{align}
and, therefore, the cubic action is now given by
\begin{align}
    \delta_3 S \simeq \int \md^3x\md t\, a^3 \alpha^2  \Big[ & \dot{\tilde\xi}^2  \tilde\xi - 2  \dot{\tilde\xi} \partial_i \tilde\xi \partial_i \partial^{-2} \dot{\tilde\xi} +  \tilde\xi \frac{(\partial_i \tilde\xi)^2}{a^2} +2 \frac{\rho_2^2}{\rho_4} \frac{\dot{\tilde\xi}}{H} \frac{(\partial_i \tilde\xi)^2}{a^2} \nonumber \\
    & +  2 \frac{\xi}{\alpha} \frac{\dot{\tilde\xi}}{H} \frac{\delta \L}{\delta \tilde\xi}\Big].
\end{align}
Therefore, inside the GLVP class there is one additional operator in comparison to the cuscuton model, namely
\begin{align}
\int \md^3x \md t\, 2 a^3  \alpha^2 \frac{\rho_2^2}{\rho_4} \frac{\dot{\tilde\xi}}{H} \frac{(\partial_i \tilde\xi)^2}{a^2}.
\end{align}
If $\rho_2=0$ we exactly recover the bispectrum of the cuscuton model at leading order similar to what happens at the power spectrum at next-to-leading order. 

The three point function of the new operator is given by
\begin{align}
    \langle \tilde\xi^3 \rangle =  \frac{ \alpha \rho_2^2}{2 \rho_4} (2\pi)^7 P_{\tilde\xi}^2 \delta^{(3)}(\vec k_1 + \vec k_2 + \vec k_3) \Pi_i \frac{1}{k_i^3} \mathcal{A}
\end{align}
where
\begin{align}
    \mathcal{A}= \frac{k_1^2}{K} (\vec k_2 \cdot \vec k_3)\left( 1 + \frac{k_2 + k_3 }{K} + \frac{2 k_2 k_3 }{K^2} \right) + \mathrm{perm.}
\end{align}
where $K=k_1+k_2+k_3$. It turns out that the three point function is peaked in the equilateral shape, $k_1=k_2=k_3$, and the $f_{\mathrm{NL}}$ for the non-Gaussianities is given by
\begin{align}
    f_{\mathrm{NL}}^{\mathrm{equilateral}} =- \frac{85}{81} \frac{\alpha \rho_2^2}{2 \rho_4}.
\end{align}
This does not come as a surpise, given that the new interaction terms is given by derivative couplings. Note, that we have assumed that $\rho_2,\,\rho_4 \sim \mathcal{O}(1)$ so that we cannot tune them arbitrarly to  enhance or suppress the bispectrum.
Similar to the original cuscuton model the non-gaussianites are slow-roll suppressed and, therefore, outside the reach of current experimental constraints \cite{Planck:2019kim}. 

\subsubsection*{Outside GLPV }

In the general case, when $b\neq 0$ and $\rho_4\neq0$ the structure of the bispectrum is very involved due to the new non-local operators which are of the form like
\begin{align}
\label{int}
    \int \md^3 \md t a^3 \alpha^2 \Big[\left( \dot{\tilde\xi}^2 + \frac{(\partial_i \tilde\xi)^2}{a^2} \right) \frac{b \rho_2  (1+ \rho_2)}{b(1+\rho_2)^2\partial^2 + 3 \rho_4 a^2 H^2} \partial^2 \tilde \xi \Big].
\end{align}
They will contain poles either on the real ($b\rho_4 <0$) or imaginary axis ($b \rho_4 >0$). We will focus on two specific cases. 

First, let us note that if $\rho_2=\rho_4=0$ we recover the bispectrum of standard cuscuton since the interaction in~\eqref{int} vanishes. This is again similar to the situation at the linear level where we have seen that considering only kinetic detuning does not lead to a modification of the power spectrum up to next-to-leading order. 

On the other hand if $\rho_4=0$ but $\rho_2\neq 0$ the cubic action simplifies to
\begin{align}
    \delta_3 S\vert_{\rho_4=0} \simeq  \int \md^3x\md t\, & a^3 \alpha^2  \Big[ \frac{1}{1+\rho_2}\tilde\xi \left( \dot{\tilde\xi}^2 +   \frac{(\partial_i \tilde\xi)^2}{a^2} \right)-  \frac{2}{1+\rho_2}  \dot{\tilde\xi} \partial_i \tilde\xi \partial_i \partial^{-2} \dot{\tilde\xi} \nonumber \\
    &   +   \frac{6 \rho_2 (\rho_2 + b + b\rho_2) }{b(1+\rho_2)^2} H \dot{\tilde\xi}  \partial_i \tilde\xi \partial_i \partial^{-2} \tilde\xi  +  2  \frac{\xi}{\alpha} \frac{\dot{\tilde\xi}}{H} \frac{\delta \L}{\delta \tilde\xi}\Big].
\end{align}
Besides the usual three standard operators but with a different overall factor there is just one additional new operator, namely
\begin{align}
    \int \md^3x \md t\, 6 a^3 \alpha^2 \frac{ \rho_2 (\rho_2 + b + b\rho_2) }{b(1+\rho_2)^2} H \dot{\tilde\xi}  \partial_i \tilde\xi \partial_i \partial^{-2} \tilde\xi.
\end{align}
The three point function for the new operator is given by
\begin{align}
    \langle \tilde\xi^3 \rangle \simeq - \frac{3 \alpha \rho_2 (\rho_2 + b +b \rho_2) }{4b(1+\rho_2)^2} (2\pi)^7 P_{\tilde\xi}^2 \delta^{(3)}( \vec k_1 + \vec k_2 + \vec k_3 ) \Pi_i \frac{1}{k_i^3} \mathcal{A}
\end{align}
where
\begin{align}
    \mathcal{A} =  \frac{k_1^2}{k_3^2} (\vec k_2 \cdot \vec k_3 ) \left(K - k_1 (\gamma_E + \log K \tau_e) - \frac{k_2 k_3 }{K} \right) + \mathrm{perm.}
\end{align}
and $\tau_e$ is the conformal time at the end of inflation. Note, that this expression diverges in the asymptotic limit $\tau_e \rightarrow 0$. However, as discussed in section \ref{sec:slow_roll_approximation} the subhorizon approximation breaks down shortly after horizon crossing. Therefore, we normalize the integral via fixing $\gamma_E + \log K \tau_e \simeq 0$. The shape of the three point function can be approximated as
\begin{align}
    \mathcal{A} \simeq  \frac{k_1^2}{k_3^2} (\vec k_2 \cdot \vec k_3 )  \left( K - \frac{k_2 k_3}{K}\right)+ \mathrm{perm.}
\end{align}
which is peaked in the squeezed limit, $k_3 \ll k_1 \sim k_2$. The $f_{\mathrm{NL}}$ in the squeezed configuration is given by
\begin{align}
    f_{\mathrm{NL}}^{\mathrm{squeezed}} = - \frac{ 10\rho_2 (\rho_2 + b + b\rho_2) }{4 b (1+\rho_2)^2}  \alpha.
\end{align}
Similar to the previous case we assumed that $1+\rho_2,\, b \sim \mathcal{O}(1) $ and, therefore, we cannot tune them arbitrarily to obtain an enhanced bispectrum. 

\section{Conclusions}
\label{sec:Conclusion}
In this work we have discussed the impact of the (generalized) cuscuton model on single scalar field inflation. For simplicity, we have focused on models which have the same background evolution as the original cuscuton model which allows for a new non-trivial power-law scenario introduced in \cite{Ito:2019fie}. In general, by integrating out the non-dynamical scalar fields we obtain a non-trivial scale dependent sound speed and friction term. However, expanding them in terms of the slow-roll parameters we have shown that it does not spoil the standard single scalar field inflationary model.  Indeed the corrections coming from the (generalized) cuscuton field are in most cases slow-roll suppressed and, therefore, the scalar power spectrum coincides up to a different normalization due to a new slow-roll parameter with the standard single scalar field result consistent with previous results in \cite{Ito:2019fie}. While for the cuscuton model the tensor spectrum is identical to the canonical scalar field the generalized models can impact the tensor spectrum by providing a modified sound speed and friction term.

By independently tuning the new slow-roll parameters it leads to a wider parameter range which could be used to reconcile ruled out models as the power-law inflation scenario. However, due to the degeneracy between the different slow-roll parameters we would need a direct detection of the tensor power spectrum in order to disentangle the effects from the cuscuton to other more conventional single scalar field models. 

Another way to look for specific signatures of (generalized) cuscuton models (w.r.t to standard single-field models of inflation) is represented by primordial non-Gaussianity. We have analyzed the bispectrum by expanding the cubic action from the start in orders of the slow-roll parameters. For generalized cuscuton models the non-gaussianities are proportional to the slow-roll parameter similar to the conventional canonical single scalar field models. Indeed, for the original cuscuton model in the case of a power-law scenario there is no difference to the usual canonical inflation model up-to-leading order in slow-roll. Therefore, they are outside the current reach of experimental constraints.  However, generalized cuscuton provide, in general, a different shape for the bispectrum which might be possible to test in future experiments.

\acknowledgments
A.G. is supported  by the grant No. UMO-2021/40/C/ST9/00015 from the National Science Centre, Poland. N.B. and S.M acknowledge support from the COSMOS network (www.cosmosnet.it) through the ASI (Italian Space Agency) Grants 2016-24-H.0 and 2016-24-H.1-2018.

\appendix

\section{Instantaneous modes}
\label{app:Instantaneous_modes}
To analyze the presence of the extra modes let us consider the linear scalar field perturbations $\varphi = \bar \varphi + \delta \varphi$ neglecting the metric perturbations. Further, we set $V=0$ since the form of the potential does not impact the result. At second order the Lagrangian is given by
\begin{align}
    \delta_2\mathcal{L} = - \frac{1}{2} \sqrt{-g} \frac{\mu^2}{\sqrt{\bar X}} \nabla^\alpha \delta \varphi \nabla^\beta \delta \varphi \left(  g_{\alpha\beta} - \frac{ \nabla_\alpha \bar\varphi \nabla_\beta \bar \varphi}{\nabla_\mu \bar\varphi \nabla^\mu \bar \varphi} \right),
\end{align}
where $X= - g^{\mu\nu} \partial_\mu \varphi \partial_\nu \varphi$.
Decomposing $g_{\alpha\beta} =   h_{\alpha\beta} - n_\alpha n_\beta $, where $n_\alpha$ is the normal vector to the hypersurface of constant time and $h_{\alpha\beta}$ is the induced metric, we can see that, in general, there is a dynamical scalar degree of freedom. However, the kinetic term of the perturbation of the scalar field $\delta\varphi$ vanish if we identify the time flow with the background evolution of the scalar field $n_\alpha =- \nabla_\alpha \bar \varphi/ \sqrt{\bar X}$. For instance, in the Minkowski spacetime $g_{\mu\nu} = \eta_{\mu\nu}$ and for a homogeneous background scalar field $\bar \varphi = \bar \varphi(t)$ the second order Lagrangian simplifies to
\begin{align}
    \delta_2 \mathcal{L} = -\frac{1}{2} \frac{\mu^2}{\vert \dot{\bar\varphi} \vert} (\partial_i \delta \varphi)^2.
\end{align}
Consequently, there is no scalar degree of freedom at linear order for a homogeneous scalar field background. At linear level the scalar degree of freedom can only be seen if the background is non-homogeneous \cite{Afshordi:2006ad,Iyonaga:2018vnu}.

At higher order the analysis becomes a bit more subtle. At cubic order the Lagrangian is given by
\begin{align}
    \delta_3 \mathcal{L} =  - \frac{1}{2} \sqrt{-g} \frac{\mu^2}{\sqrt{\bar X}^3} \nabla_\alpha \delta\varphi \nabla_\beta \delta\varphi \nabla_\gamma\delta\varphi \nabla^\gamma \bar\varphi \left( g^{\alpha\beta}  - \frac{\nabla^\alpha \bar\varphi \nabla^\beta \bar \varphi }{\nabla_\mu \bar \varphi \nabla^\mu \bar \varphi} \right).
\end{align}
Even identifying the time flow with the background scalar field does not cancel all the time derivatives. Indeed, considering the Minkowski background with a time-dependent background scalar field it yields
\begin{align}
    \delta_3 \mathcal{L} = \frac{1}{2} \mu^2 \frac{\dot{\bar\varphi}}{\vert \dot{\bar\varphi} \vert^3} \delta \dot\varphi ( \partial_i \delta \varphi)^2.
\end{align}
Such a term leads to an EOM, which is first order in time-derivatives, resulting in one-half degree of freedom. Similar at quartic order in Minkowski space we obtain
\begin{align}
    \delta_4 \mathcal{L} = - \frac{1}{8} \frac{\mu^2}{\vert \dot{\bar\varphi}\vert^3} (\partial_i \delta \varphi)^2 \left( 4 \delta \dot \varphi^2 + ( \partial_i\delta \varphi )^2 \right).
\end{align}
Consequently, going up to quartic order the cuscuton degree of freedom seems to be completely revived. 

The analysis can be straightforwardly extended to include the metric perturbations. In the unitary gauge $\varphi=\bar \varphi(t)$ the instantaneous mode is absent at cubic order in perturbation theory. Outside the unitary gauge but with a homogeneous background $\varphi=\bar \varphi(t) + \delta \varphi(t,x^i)$ the analysis is more subtle due to the presence of time derivative terms at higher order similar to the previous discussion in Minkowski spacetime. 

However, solving the EOM order by order the cuscuton field can be obtained perturbatively without the need of extra initial conditions \cite{Afshordi:2009tt}. 

\if{}
\subsection{Full scalar sector}
So far, we focused only on the cuscuton field neglecting the gravity contributions. As a next step let us analyze the full scalar sector of the cuscuton coupled to gravity
\begin{align}
    S = \int \md^4x\,\sqrt{-g} \left(\frac{1}{2}R +  \mu^2 \sqrt{- g^{\mu\nu} \partial_\mu \varphi \partial_\nu \varphi} - V(\varphi) \right)
\end{align}
At the quadratic level, it is straightforward to see that again there is no dynamics and the choice of the gauge has no impact. To illustrate it we present the the second order action in the unitary and flat gauge around the FLRW background. In the unitary gauge, $\delta \varphi=0$, $h_{ij}= a^2 e^{2\xi}\delta_{ij}$, $N_i = \partial_i \psi$ the second order action can be written as
\begin{align}
    \delta_2 S_{\mathrm{unitary}} = \int \md^4x\,a^3 \Big[& - 3 \dot \xi^2 + \frac{(\partial_k \xi)^2}{a^2} - 3 H^2 \delta N^2 - 2 H \delta N \frac{\partial^2 \psi}{a^2} + 2 \dot \xi \frac{\partial^2 \psi}{a^2} \nonumber \\ &  + 6 H \delta N \dot \xi - 2 \delta N \frac{\partial^2 \xi}{a^2} \Big] 
\end{align}
Solving the Hamiltonian and momentum constraint
\begin{align}
\label{eq:delta_N_unitary}
    \delta N = \frac{\dot \xi}{H} , \qquad \psi = - \frac{1}{H} \xi
\end{align}
and replacing it back into the action, it is immediate that there is no dynamic left consistent with our previous results.

Similar in the flat gauge
\begin{align}
    \delta_2 S_{\mathrm{flat}}= \int \md^4x\,a^3 \big[& - 3 H^2 \delta N^2 - 2 H \delta N \frac{\partial^2 \psi}{a^2} + \mu^2 \frac{\partial^2 \psi}{a^2} \delta \varphi - \frac{\mu^2}{2 \vert \dot \varphi \vert} \frac{(\partial_i \delta \varphi)^2}{a^2} \nonumber \\
    & + V_\varphi \delta N \delta \varphi + \frac{1}{2} V_{\varphi\varphi} \delta \varphi^2 \big]
\end{align}
Even without using the momentum and hamiltonian constraint
\begin{align}
\label{eq:delta_N_flat}
    \delta N = \frac{\mu^2}{2 H} \delta \varphi, \qquad \frac{\partial^2 \psi}{a^2} = \frac{1}{2} \left( - 3 \mu^2  + \frac{V_\varphi}{H}\right) \delta \varphi
\end{align}
it is immediate that there is no dynamics. The choice of gauge does not impact this result.

Let us now go to the cubic order. Consider first the unitary gauge we obtain
\begin{align}
    \delta_3 S = \int \md^4x\,a^3 \big[ & 3 H^2 \delta N^3 - 9 H^2 \delta N^2 \xi - 6 H \delta N^2 \dot \xi + 18 H \dot \xi \xi \delta N + 3 \dot \xi^2 \delta N - 9 \dot\xi^2 \xi \nonumber \\
    &+ 2 \frac{\partial^2 \psi}{a^2} \left( \delta N -\xi \right) ( H \delta N -\dot \xi) + 2 \frac{\partial_i \xi \partial^i \psi}{a^2} ( \dot \xi - H \delta N) + \delta N \big( - 2 \xi \frac{\partial^2 \xi}{a^2} \nonumber \\ 
    & - \frac{(\partial_i \xi)^2}{a^2}\big) + \frac{(\partial_i \xi)^2}{a^2} \xi - 4 \frac{\partial^2 \psi}{a^2} \frac{\partial_i \psi \partial^i \xi}{a^2} + \left( \frac{\partial^2\psi}{a^2} \right)^2 ( - 3\xi + \delta N) \nonumber \\ 
    & - \frac{\psi_{ij} \psi^{ij}}{a^2} ( - 3 \xi +\delta N) \big]
\end{align}
Using the solution for $\delta N$ and $\psi$ \eqref{eq:delta_N_unitary} and plug it back into the action, we can see that all the terms with time derivatives are cancelled. Since in the unitary gauge the scalar field is homogeneous this result is consistent with the full non-linear result [..].

In the flat gauge it is sufficient to consider only the cuscuton contribution since $\delta N$ and $\psi$ does not contain time derivative terms \eqref{eq:delta_N_flat}.
So we obtain {\color{blue} Test it again}
\begin{align}
    \delta_3 S_{\varphi} = \int \md^4x\,a^3 \big[& - 4 \mu^2 \frac{\dot \varphi}{\vert \dot \varphi \vert} \delta N^2 \delta \dot \varphi +  \frac{\mu^2 \dot \varphi}{2 \vert \dot \varphi\vert^3} \delta \dot \varphi \frac{(\partial_i \delta \varphi)^2}{a^2} + \mu^2 \frac{\vert\dot \varphi \vert}{2} \delta N^3 - \frac{\mu^2}{\vert \dot \varphi \vert} \delta N \frac{(\partial_i \delta \varphi)^2}{a^2} \nonumber \\
    & - \frac{1}{2} V_{\varphi\varphi} \delta \varphi^2 \delta N - \frac{1}{6} V_{\varphi\varphi\varphi} \delta \varphi^3\big].
\end{align}
After replacing $\delta N$ and partial integrating, the critical term is the same as in Minkowski space, which leads to one half degree of freedom. Therefore, we can deduce that at cubic order the different choices of gauge can lead to different results. This is in agreement with the discussion in [] that fixing the unitary gauge is not just a gauge choice but leads to additional constraints which remove the cuscuton degree of freedom. Therefore, we have to be careful in choosing the foliation of spacetime if we perform higher order calculations in contrast to the quadratic action where the choice of the gauge has no impact.

\fi

\section{Linear perturbations of generalized Cuscuton}
\label{app:Linear_perturbations_generalized_Cuscuton}
Setting $b_1=0$ and $b_2$, $\rho_2$ and $\rho_4$ to constants in time the EOM for $\delta N$ and $\psi$ from \eqref{eq:Linear_Perturbations_Generalized_Cuscuton} leads to
\begin{align}
    \delta N =& \frac{\dot \xi}{H} + \frac{\dot \chi}{2 H} \delta \chi + \frac{b }{3 (1+b)} \frac{k^2}{a^2} \psi \label{eq:delta_N_GC}, \\
    \frac{k^2}{a^2} \psi =& - \frac{3 (1+b)}{2 H^2 (6 + 2 b \alpha)} \Big( - 2 \sigma H^2 \dot \chi \delta \chi + 4 \alpha H^2  \dot\xi  -2 H \dot \chi \delta \dot \chi + 4 (1+\rho_2) H \frac{k^2}{a^2} \xi  \Big). \label{eq:psi_GC}
\end{align}
Introducing $\tilde \xi$ as
\begin{align}
    \tilde \xi = \xi - \frac{H}{\dot \chi} \delta \chi
\end{align}
and reinserting it back into the action \eqref{eq:Linear_Perturbations_Generalized_Cuscuton} the inflaton becomes non-dynamical. Solving its EOM we obtain
\begin{align}
    \delta \chi =& \sqrt{2 \alpha } \frac{b (1+\rho_2)^2 k^4 - k^2 \mathcal{H}^2 \left(-3 (\epsilon (1+\rho_2) - \rho_4) + \alpha (3 (1+\rho_2) + b \rho_4 ) \right)}{- b (1+\rho_2)^2 k^4 + d_3 \mathcal{H}^2 k^2 + d_4 \mathcal{H}^4} \tilde\xi \nonumber \\
    & + \sqrt{2 \alpha^3} \mathcal{H} \frac{b(1+\rho_2) k^2 + 3\sigma \mathcal{H}^2 }{- b (1+\rho_2)^2 k^4 + d_3 \mathcal{H}^2 k^2 + d_4 \mathcal{H}^4} {\tilde \xi}^{\prime} \label{eq:delta_chi_GC}
\end{align}
where 
\begin{align}
    d_3 =& - b \alpha^2 - 3 (\epsilon (1+\rho_2) - \rho_4) + \alpha (3 +6\rho_2 + b(\epsilon(1+\rho_2) +\rho_4)), \\
    d_4 =& - 3 \alpha \sigma ( 3 - \sigma + b \alpha ).
\end{align}
Substituting it back into the action we obtain
\begin{align}
    \delta_2 S= \int \md^3k \md \tau\,\frac{1}{2} z^2(k,\tau) \Big[ \tilde\xi^{\prime 2} - c_s^2(k,\tau) \frac{k^2}{a^2} \tilde \xi^2 \Big]
\end{align}
where
\begin{align}
     z^2(k,\tau) =& 2 a^2 \alpha \frac{  b (1+\rho_2)^2 k^4 + d_1 \mathcal{H}^2 k^2 + d_2 \mathcal{H}^4 }{ b (1+\rho_2)^2 k^4 + d_3 \mathcal{H}^2 k^2 + d_4 \mathcal{H}^4 }, \\
    c_s^2(k,\tau) =& \frac{b^2 (1+\rho_2)^4 k^8 + b d_5 \mathcal{H}^2 k^6 + d_6 \mathcal{H}^4 k^4 + d_7 \mathcal{H}^6 k^2 + d_8 \mathcal{H}^8}{b^2(1+\rho_2)^2 k^8 + b d_9 \mathcal{H}^2 k^6 + d_{10} \mathcal{H}^4 k^4 + d_{11} \mathcal{H}^6 k^2 + d_{12}  \mathcal{H}^8}
\end{align}
with 
\begin{align}
    d_1 =& 3 ( \alpha + 2\alpha \rho_2 - \epsilon (1+\rho_2) +\rho_4), \\
    d_2 =& - 9 \sigma \alpha, \\
    d_5=& (1+\rho_2)^2 \big( b \alpha^2 -2 b \epsilon^2 (1+\rho_2)^2 + 6 \epsilon (1+\rho_2) (1+b + b \rho_2) - 3 \alpha ( 2 + 4 \rho_2 + \rho_2^2) \nonumber \\
    &+ b \alpha (- 6 + \epsilon (1+\rho_2) - 9 \rho_2 -3  \rho_2^2 - 2 \rho_4 ) - 6 \rho_4 \big) \\
    d_6=& 3 (3+ b\alpha) (\alpha \rho_2 + \rho_4)^2 + 3 (1+\rho_2) \sigma \big(6 b \alpha (1+\rho_2) +b^2 \alpha^2 (1+\rho_2)- 6 (\alpha \rho_2 +\rho_4)\big) \nonumber \\
    & -3 (-3 + 2 b \alpha) (1+\rho_2)^2 \sigma^2, 
\end{align}
\begin{align}
    d_7 =& - 9 \alpha \sigma \big( 2 (3+b\alpha) (\alpha \rho_2 + \rho_4) - (6 + 6 \alpha+ 6\rho_2 + \rho_4) \sigma + (1+\rho_2) \sigma^2 \big), \\
    d_8 =& 27 \alpha^2 \sigma^2 (3+b\alpha - \sigma), \\
    d_9 =& (1+\rho_2)^2 \big( - b \alpha^2 \rho_2 (3+2\rho_2) - 6 \rho_4 + 2 (1+\rho_2) \sigma (3+ b (1+\rho_2) (3-\sigma) ) \nonumber \\
    & - \alpha (b (2\rho_4 +3 \sigma ) + \rho_2^2 (3-3b+ 4 b\sigma )+ \rho_2 (6-3 b + 7 b \sigma ) )  \big), \\
    d_{10}=& 9 \rho_4^2 + b \rho_2 \alpha^3 ( (3-b) \rho_2 (1+\rho_2) + b \rho_4 ) + 6 (1+\rho_2) \rho_4  \sigma (-3 + 2b (1+\rho_2) (-3 +\sigma)) \nonumber \\
    & + 3(1+\rho_2)^2 \sigma (3-4b (1+\rho_2) (-3+\sigma)) + \alpha \big( - 3\rho_2 \rho_4 (-6 + b - (3-b) \rho_2) \nonumber \\
    & + 6 b \rho_4^2 - (1+\rho_2) \big( 9 \rho_2 (2+\rho_2) - 9 b \rho_2 (2+\rho_4) - 6 b (3+\rho_4) + b^2 (1+\rho_2) \nonumber \\
    &\times (15+15\rho_2 + 7 \rho_4 ) \big) \sigma + 2 b(1+\rho_2)^2 (-9 - 6 \rho_2 + 4 b(1+\rho_2))\sigma^2 \big) \nonumber \\
    & + \alpha^2 \big( b \rho_2 (\rho_4 (9 + b (-1 +\sigma )) + 37 b \sigma ) + \rho_2^2 ( (3-b) (3+b\rho_4) + b (3 + 29 b) \sigma) \nonumber \\
    & + b^2 ( \rho_4^2 + (15 + \rho_4) \sigma ) + \rho_2^3 (9 + b ( -3 + \sigma (3+7b)  ))\big), \\
    d_{11}=& - 3 \sigma \big( b^2 \alpha^2 (-2\alpha \rho_2 (1+\rho_2) + \alpha \rho_4 (3+\rho_2) - \rho_4 ( 3 + 3 \rho_2 + \rho_4) + (1+\rho_2) \sigma \nonumber \\
    &\times (-5 - 5 \rho_2 +\rho_4)  ) + 3 \alpha (-6 +3 \rho_2 - \rho_4 +3 \sigma ) (-\rho_4 + \sigma + \sigma \rho_2 ) - 6 (3-\sigma) \nonumber \\
    &\times (-\rho_4 +\sigma + \sigma \rho_2  )^2 + 3 \alpha^2 \rho_2  (6 + 6 \rho_2 + \rho_4 - 3(1+\rho_2) \sigma ) + b \alpha ( \alpha^2 \rho_2 \nonumber \\
    & \times (6 + 6\rho_2 + \rho_4) + 9 (1+\rho_2) \rho_4 (-1 +\sigma) + (1+\rho_2)^2 (3-4\sigma) \sigma + \rho_4^2 (-9 + 2 \sigma) \nonumber \\
    & + \alpha ( -6\rho_2 (1+\rho_2) + \rho_4 (15 + \rho_4) -12 \sigma - 4 \rho_2 (4 + \rho_2) \sigma + (-2 +\rho_2 ) \rho_4 \sigma  )  )  \big), \\
    d_{12}=& - 9\alpha (3 + b\alpha -\sigma) \sigma^2 \big( (1+b) \alpha (1+\rho_2 +\rho_4) - 4 (\alpha - \rho_4 + \sigma + \epsilon \rho_2 ) \big).
\end{align}

\bibliography{bibliography}
\bibliographystyle{JHEP}

\end{document}